\documentclass[
floatfix, 
jcp,
aip, 
a4paper,
preprint
]{revtex4-1}

\pdfoutput=1

\usepackage{graphicx}
\usepackage{amssymb}
\usepackage{amsmath}
\usepackage[utf8]{inputenc}
\usepackage[T1]{fontenc}
\usepackage{cancel}
\usepackage[usenames,dvipsnames]{color}
\usepackage{psfrag}
\usepackage{epsfig}
\usepackage{bbm}
\usepackage{bm}
\usepackage{dsfont}
\usepackage{soul}
\usepackage{colonequals}
\usepackage{float}
\usepackage{enumitem}

\usepackage[caption=false]{subfig}

\newcommand{\bs}[1]{\boldsymbol{#1}} 
\newcommand{\divergence}{\nabla \cdot} 
\newcommand{\makeEffectiveAverageDensitySymbol}{\bar{\rho}_{\text{eff}}} 


\begin{document}

\title{Freezing of a soft-core fluid in a one-dimensional potential:\\
Predictions based on a pressure-balance equation
} 
\author{Alexander Kraft}
\email{alexander.kraft@tu-berlin.de}
\author{Sabine H.~L.~Klapp}
\email{sabine.klapp@tu-berlin.de}

\affiliation{
  Institut f\"ur Theoretische Physik,
  Hardenbergstr.~36,
  Technische Universit\"at Berlin,
  D-10623 Berlin,
  Germany}
  
\date{\today}

\begin{abstract}
Using concepts from classical density functional theory (DFT) we investigate the freezing of a two-dimensional (2D) system of ultra-soft particles
in a one-dimensional (1D) external potential; a phenomenon often called laser-induced freezing (LIF). In the first part of the paper, we present numerical results
from free minimization of a mean-field density functional for a system of particles interacting via the \mbox{GEM-4} potential. We show that the system does indeed display a LIF transition, although the interaction potential is markedly different from the cases studied before. We also show that 
one may consider the (suitably defined) effective density within the potential wells, $\bar{\rho}_{\text{eff}}$, as a control parameter of LIF, rather than the amplitude
of the external potential as in the common LIF scenario. In the second part, we suggest a new theoretical description of the onset of LIF which bases on the pressure balance equation relating the pressure tensor and the external potential. Evaluating this equation  
for the modulated liquid phase at effective density $\bar{\rho}_{\text{eff}}$ and combining it with the (known) stability threshold of the corresponding bulk fluid, we can predict the critical effective density or, equivalently, the potential
amplitude related to the onset of LIF. Our 
approach yields very good results for the model at hand, and it is transferable, in principle, to other model systems.
\end{abstract}

\maketitle

\section{Introduction \label{SEC:INTRO}}
Laser-induced freezing of a two-dimensional (2D) colloidal system describes the intriguing phenomenon in which a 1D standing-wave pattern of interfering laser beams induces a liquid-solid freezing transition, which displays density modes other than those directly excited. This phenomenon provides an excellent example of how the equilibrium structure and diffusion of colloidal systems can be manipulated by a periodic potential, see Refs.~\onlinecite{Bechinger2002, Bechinger2007, Jenkins2008} for reviews. 
LIF was first discovered experimentally by Chowdhury, Ackerson, and Clark~\cite{Chowdhury1985} who investigated a 2D monolayer of charged spherical particles subjected to a 1D periodic light field.
Provided that the wavelength of the light field is commensurate with the mean particle distance, the modulated liquid (characterized by 1D symmetry breaking) appearing at low potential amplitudes freezes into a structure with quasi-long range positional
order in both directions. This observation inspired a considerable amount of investigations by theory~\cite{Chakrabarti1994, Das1998, Das1999a, Frey1999, Radzihovsky2001, Rasmussen2002, Chaudhuri2004, Nielaba2004, Chaudhuri2006, Luo2009}, computer simulations~\cite{Loudiyi1992b, Chakrabarti1995, Das1999a, Das1999b, Das2001, Strepp2001, Strepp2002, Strepp2003, Chaudhuri2004,Chaudhuri2005, Chaudhuri2006, Burzle2007, Luo2009} and experiments~\cite{Loudiyi1992a, Wei1998, Bechinger2000, Bechinger2001}. Major points of discussion concerned the order of the LIF freezing transition, as well as the origin of the re-entrant melting experimentally observed at high laser intensities~\cite{Wei1998}. Indeed, phenomenological approaches like the Alexander-McTague theory~\cite{Alexander_McTague1978, Chowdhury1985}, which foots on a Ginzburg-Landau free energy, turned out to be incapable of
describing these issues due to the negligence of fluctuations. A major step towards an understanding of the full LIF scenario was provided by Frey, Nelson, and Radzihovsky\cite{Frey1999, Radzihovsky2001}, who used
the concept of dislocation-mediated melting described by KTHNY theory~\cite{Kosterlitz1973, Halperin1978, Nelson1979, Young1979}. Their results were later confirmed
by extensive numerical (Monte-Carlo) simulation studies~\cite{Strepp2001, Strepp2002, Strepp2003, Burzle2007}.

However, whereas the physical concepts underlying LIF are well settled for more than two decades, {\em quantitative} theoretical predictions for LIF in different model systems (i.e., different interaction potentials)
remain to be difficult. This is one of the main topics of the present paper.
Indeed, the need for improved quantitative predictions is becoming relevant again in view of recent experimental studies showing, e.g., that periodic (light) potentials can also cause
other transitions such as demixing \cite{Capellmann2018}, or for investigating the ordering of {\em soft} (e.g., polymer-grafted) particles in periodic potentials \cite{Schoch_SoftMatter2014, Schoch_Langmuir2014}. The behaviour of soft particles
is the second theme of our work.

The most established microscopic approach to the freezing of liquids is classical density functional theory (DFT). Here, the key quantity is a grand-canonical free energy functional, which directly involves
the particle interactions and is minimized by the
equilibrium density profile. A particularly prominent DFT approach is the Ramakrishnan-Yussouff (RY) theory\cite{RamakrishnanYussouff1979,Ramakrishnan1982}. Here, the interaction part of the free energy of the solid phase is functionally expanded up to second order in the density
around the liquid ("reference") phase, yielding the direct (two-particle) correlation function of the liquid as a key input. The minimization is then performed by using a suitable {\em ansatz} for the density in the solid phase
(typically, the lowest-order Fourier components). Comparing the grand potentials of the solid and liquid phase finally allows to locate the transition. However, while RY theory has shown to be quite successful in 
predicting the freezing of a large variety of bulk systems (see Ref.~\onlinecite{Singh1991} for a review), including systems with anisotropic interactions~\cite{Klapp1997, Klapp1998, Klapp2000}, the application to LIF is less straightforward since here already the reference state, that is, the modulated liquid, is inhomogeneous. Still, there are several investigations applying concepts of RY theory to LIF (see, e.g., Refs.~\onlinecite{Chakrabarti1994} and \onlinecite{Luo2009}).

In the present paper we propose an alternative approach, which is based on DFT but uses the (exact) pressure balance equation\cite{Evans1979} as the main ingredient. This equation, which expresses hydrostatic equilibrium, relates the divergence of the stress (i.e., the negative pressure) tensor of the inhomogeneous liquid to the force generated by the external potential. We evaluate the pressure balance equation for the modulated liquid phase, using a parametrized ansatz for the density profile which involves the {\em effective} density inside the potential wells, $\bar{\rho}_{\text{eff}}$.
With this we obtain an effective-fluid equation involving the isotropic part of the pressure tensor, a deviatoric contribution stemming from the inhomogeneities, and the external potential. The resulting equation is then combined with the stability threshold of the bulk system against freezing, which is assumed to be known. This finally enables us to make a quantitative prediction for the onset of LIF.

We here apply our approach to a 2D system of "ultra-soft" particles interacting via a generalized Gaussian, specifically the GEM-4 potential\cite{Mladek2005}, whose freezing behaviour in the absence of a potential is well understood \cite{Mladek2006, Mladek2007, Archer2014, Prestipino2014}. 
We note in this context that the freezing of ultra-soft particles displays markedly different features as compared to conventional fluids with a strongly repulsive core, such as cluster crystallization\cite{Likos1998, Likos2001, Mladek2005, Mladek2006, Mladek2007, Archer2014, Prestipino2014}.
The freezing behaviour of such particles in presence of a 1D potential has not been studied so far (in contrast to other phenomena induced by a 1D potential such as freezing in a slit-pore\cite{VanTeeffelen2009} and magnetic pattern
formation \cite{Lichtner2014}).
As an external potential we here consider both, a cosine potential in analogy to earlier studies of LIF, and a periodic potential based on Gaussian functions. In both cases we focus on a commensurate situation.
The soft character of the two-particle interaction allows for a mean-field-like treatment of the excess part of the free energy.
However, our strategy to predict LIF based on the pressure equation can also be transferred to other models.

The remainder of the article is structured as follows: 
In Sec.~\ref{SEC:Model_and_density_functional_theory}, we introduce our 2D model system of ultra-soft particles and the two types of 1D periodic substrates, as well as the corresponding density functional
in mean-field approximation. In Sec.~\ref{SEC:Free_minimization_within_DFT}, we present results from a (numerical) "free" minimization for various average densities and substrate potentials. In this way we demonstrate
that LIF indeed occurs for the ultra-soft system at hand. By studying different variants of the external potential, we also propose that LIF can be understood as a density-driven transition controlled
by the effective density inside the potential wells. The theoretical approach to predict the onset of LIF is outlined in Sec.~\ref{SEC:Towards_a_theoretical_description}, where we consider an integrated form of the (exact) 
pressure balance equation. Explicit calculations for the present model system are described in Sec.~\ref{SEC:Explicit_calculations_for_our_model_system}, where we compare different variants of the theoretical description
with the results from free minimization. 
We conclude and give an outline for future research in Sec.~\ref{SEC:Conclusion}.

\section{Model and density functional theory \label{SEC:Model_and_density_functional_theory}}

\subsection{Model system}
\begin{figure}
    \centering
    \includegraphics[width=0.45\textwidth]{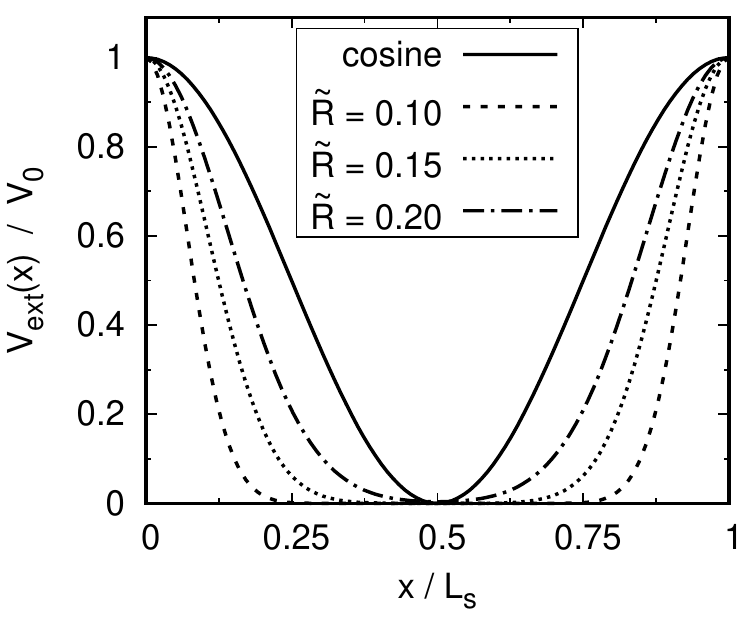}
    \caption{Comparison of the cosine potential (shifted by $V_0/2$) and the Gaussian substrate potential [see Eqs.~\eqref{Eq_cosine_substrate} and \eqref{Eq_Gaussian_substrate}] for the same values of $V_0$ and $L_s$, and different (dimensionless) ranges of the Gaussian peak, $\tilde{R} = R_{g}/L_{s}$.
	An increase of $\tilde{R}$ of the Gaussian substrate yields an increase of the energetic cost for positional deviations from the location of the minimum.
    }
    \label{Fig_gaussian_for_different_R_Peak}
\end{figure}

We consider a 2D colloidal system (located on the $x-y$ plane of the coordinate system) subjected to two variants of 1D periodic substrate potentials.
The simplest variant is the harmonic (cosine) substrate potential 
\begin{align}
	    V_{\text{ext}}(\bs{r} ) =  \frac{V_0}{2} \cos\left( \frac{2 \pi}{L_{s}} x \right),
	    \label{Eq_cosine_substrate}
\end{align}
with periodicity $L_{s}$ and amplitude~$V_0$~(the factor 1/2 was introduced such that $V_0$ denotes the potential difference between potential maxima and minima), and the position vector $\bs{r} = (x,y) \in \mathbb{R}^2$.
This functional form has also been used in earlier theoretical\cite{Chakrabarti1994, Das1998, Das1999a, Frey1999, Radzihovsky2001, Rasmussen2002, Chaudhuri2004, Nielaba2004, Chaudhuri2006, Luo2009} and simulation studies\cite{Loudiyi1992b, Chakrabarti1995, Das1999a, Das1999b, Das2001, Strepp2001, Strepp2002, Strepp2003, Chaudhuri2004,Chaudhuri2005, Chaudhuri2006, Burzle2007, Luo2009} of LIF and was realized in experiments of charge stabilized polystyrene spheres subjected to a 1D periodic light field by Bechinger \textit{et al.}~\cite{Bechinger2001}.
The second variant is an artificial \emph{ansatz} for the substrate potential whose main advantage is its tunability: It allows to independently adjust the amplitude~$V_0$ \emph{and} the available space in the vicinity of the minima. 
Specifically, we consider the Gaussian substrate
\begin{align}
	 V_{\text{ext}}(\bs{r} ) = \sum_{m \in \mathbb{Z}}  V_0 \exp\left( - \left( \frac{ x-m {L_{s}} }{ R_{g}} \right)^2 \right),
	 \label{Eq_Gaussian_substrate}
\end{align}
where $R_{g}$ is a measure of the range of the Gaussian.
We compare the cosine substrate with the Gaussian substrate in Fig.~\ref{Fig_gaussian_for_different_R_Peak}, which illustrates the tunability of the Gaussian ansatz: 
For fixed periodicity~$L_s$ and amplitude~$V_0$, increasing $R_{g}$ increases the energetic cost for deviations of the particle position from the exact location of the minimum.
This effectively reduces the available space around the minimum.
Although not denoted as such in Ref.~\onlinecite{Zaidouny2013}, from Fig.~7(a) in Ref.~\onlinecite{Zaidouny2013} it seems that the authors have experimentally realized a Gaussian substrate by a scanning optical line tweezer.

To study the influence of the 1D periodic potentials introduced above, we consider (for reasons outlined below) a 2D system of ultra-soft particles, whose interaction is given by the generalized exponential model of index $n$ (GEM-$n$),
\begin{align}
	\label{Eq_Interaction_Potential}
    V( |\bs{r}_1 - \bs{r}_2| ) = \epsilon \, \exp\left( - \left(\frac{  |\bs{r}_1 - \bs{r}_2|  }{R}\right)^n   \right).
\end{align}
In Eq.~\eqref{Eq_Interaction_Potential}, $\bs{r}_1$ and $\bs{r}_2$ denote the particle positions, $\epsilon$~denotes the interaction strength, and $R$ denotes the range of the interaction.
This model interpolates~\cite{Mladek2005} between the Gaussian core model (GCM, $n=2$), first introduced by Stillinger~\cite{Stillinger1976}, and the penetrable sphere model (PSM) introduced by Likos \textit{et al.}~\cite{Likos1998} ($n\to\infty$).

According to a criterion stated by Likos \textit{et al.}~\cite{Likos2001PhysRevE}, particles interacting via GEM-$n$ potentials with $n > 2$ tend to build clusters where they "sit on top of each other"~\cite{Likos1998}, as in the PSM.
Upon increase of the density, these clusters freeze into cluster crystals with density-\textit{independent} lattice constant.
This is achieved by multiple occupation of lattice sites~\cite{Likos1998, Likos2001PhysRevE}.
Indeed, the magnitude of the 2D reciprocal lattice vector of the first shell (being  $4 \pi / (\sqrt{3} a)$ where $a$ is the lattice constant) is essentially given by the value of the wave number  $k_*$, for which the Fourier transform of the interaction potential $\tilde{V}(k)$ has its negative minimum\cite{Likos2007}. 
It will later turn out that this feature is an advantage for the present analysis.

In this work, we are especially interested in the \mbox{GEM-4} model.
First, the interaction potential is continuous as opposed to the discontinuous PSM model.
Second, its phase diagram in the absence of external potentials is well understood in both, 3D~\cite{Mladek2006, Mladek2007} and in 2D~\cite{Archer2014, Prestipino2014}, including the cluster crystallization with density-independent lattice constant.
Throughout this work, we thus fix $n=4$ in Eq.~\eqref{Eq_Interaction_Potential}, and denote all length scales in units of $R$, the range of the particle interaction. 
Unless stated otherwise, the particle interaction strength is set to $\beta \epsilon = 1$, where $\beta = 1/ k_B T$ (with $k_B$ being Boltzmann's constant and T being the temperature).

Our reasoning to consider ultra-soft particles to study LIF is threefold: 
(i)~As will be demonstrated later, the \mbox{GEM-4} model displays a LIF transition although the cluster crystallization mechanism is fundamentally different from the crystallization of charged spheres or hard discs for which LIF was studied before\cite{Chakrabarti1994, Das1998, Das1999a, Frey1999, Radzihovsky2001, Rasmussen2002, Chaudhuri2004, Nielaba2004, Chaudhuri2006, Luo2009, Loudiyi1992b, Chakrabarti1995, Das1999a, Das1999b, Das2001, Strepp2001, Strepp2002, Strepp2003, Chaudhuri2004,Chaudhuri2005, Chaudhuri2006, Burzle2007, Luo2009, Loudiyi1992a, Wei1998, Bechinger2000, Bechinger2001}.
We note in this context that the (negative) minimum of $\tilde{V}(k)$ for the GEM-4 model in 2D occurs at $k_* R \approx 5.1$,  yielding a lattice constant of $a/R \approx 1.4$.
Indeed, we found this value in DFT calculations of the bulk system.
(ii)~We expect that due to the density-independent lattice constant characterizing the solid state, 
the usual competition between the lattice constant formed by the crystallizing fluid, on the one hand, and the substrate periodicity $L_s$ on the other hand, is less severe.
For a fixed substrate periodicity $L_s$, this simplifies the study of LIF at different densities~$\bar{\rho}$.
(iii)~The third motivation is the numerical ease of the treatment of the system via DFT. 
This allows us to systematically scan large portions of the phase diagram.

Whereas effective interactions of GEM-2 (i.e. Gaussian) type are frequently observed\cite{Dautenhahn1994, Louis2000PhysRevLett, Bolhuis2001, Likos2002, Goetze2004} (also see Refs.~\onlinecite{Likos2001, Ballauff2004} for a review), explicit realizations of particles that possess GEM-$n$ effective interaction with $n>2$ and show the clustering property, are scarce.
In computer simulations of suitably designed amphiphilic dendrimers, such potentials were obtained as coarse-grained interaction potentials\cite{Mladek2008} between the centrers of masses, and the existence of cluster crystals was indeed demonstrated\cite{Lenz2012}.
However, the  cluster crystals have not yet been observed in real experiments.

Nonetheless, the GEM-4 model is convenient to study more fundamental questions, such as the quantitative influence of periodic substrates on freezing.

\subsection{Density functional theory \label{SUBSEC:Model_and_density_functional_theory_subsection_DFT}}
For inhomogeneous systems in thermal equilibrium, the central quantity of interest is the one-body density distribution, $\rho(\bs{r})$.
We calculate this quantity using DFT\mbox{\cite{Evans1979,Evans1992}}.
This is the state-of-the-art microscopic theory to describe both, the fluid and the crystal phase, within the same theoretical framework.

The key idea of DFT is that the equilibrium density profile, $\rho_{\text{eq}}(\bs{r})$, minimizes the grand potential functional 
\begin{align}
 \Omega[\rho] &= F[\rho] +  \int d\bs{r} \rho(\bs{r}) V_{\text{ext}}(\bs{r})  - \mu \int d\bs{r} \rho(\bs{r}) 
 \label{Eq_grand_potential_functional}
\end{align}
with chemical potential~$\mu$, external potential~$V_{\text{ext}}(\bs{r})$, and the intrinsic Helmholtz free energy functional \mbox{$F[\rho] = F_{\text{id}}[\rho] + F_{\text{exc}}[\rho]$}.
The ideal gas contribution of $ F[\rho]$ is known exactly,
\begin{subequations}
	\begin{align}
		F_{\text{id}}[\rho] &= k_B T \int d\bs{r} \rho(\bs{r}) \left[ \ln(\Lambda^2 \rho(\bs{r}))-1 \right], 
		\label{Eq_ideal_gas_free_energy_functional} \\
\intertext{where $\Lambda$ is the de Broglie wavelength.
The excess free energy~$F_{\text{exc}}$, which describes the impact of the interactions between particles, has to be approximated for most types of interactions.
Here, we employ the mean-field (MF) approximation for $F_{\text{exc}}$ that is well established for the description of ultra-soft particles at high density\cite{Likos2001},}
		F_{\text{exc}}[\rho] &=  \frac{1}{2} \int d\bs{r} \int d\bs{r}' \big[ \rho(\bs{r})  V(\bs{r}-\bs{r}')  \rho(\bs{r}') \big].
	\label{Eq_excess_free_energy_functional}
	\end{align}
	\label{Eq_intrinsic_free_energy_functional}
\end{subequations}
For the homogeneous fluid phase, the high accuracy of the mean-field approximation for different ultra-soft particles was frequently demonstrated.
Applications include the Gaussian core model~\cite{Lang2000, Louis2000PhysRevE}, 
mixtures thereof~\cite{Archer2004}, 
and the GEM-$n$ model~\cite{Likos2007, Archer2014}, especially $n=4$.
For inhomogeneous phases, such as cluster crystals, the MF-DFT is further supported by agreement with Monte-Carlo simulation data for GEM-4 particles~\cite{Mladek2006, Mladek2007}.
Furthermore, the validity of the above excess free energy~$F_{\text{exc}}$ for bounded interaction potentials was also proven for arbitrary inhomogeneous phases~\cite{Likos2007}.

Apart from the direct connection to particle interactions, a further major benefit of the DFT treatment is the possibility of a free (numerical) minimization in which no \textit{a priori} information of the spatial form of $\rho(\bs{r})$ is assumed. 
The fact that the equilibrium density~$\rho_{\text{eq}}(\bs{r})$ minimizes the grand potential functional~$\Omega[\rho]$ implies that $\delta \Omega[\rho]/\delta \rho(\bs{r}) |_{\rho_{\text{eq}}} = 0$.
This results in the Euler-Lagrange equation,
\begin{align}
    \rho	(\bs{r})=\Lambda^{-2} 
    \exp\left[\beta \mu  -\beta V_{\text{ext}}(\bs{r}) - \beta \frac{\delta F_{\text{exc}}[\rho]}{\delta \rho(\bs{r})} \right]
    \label{Eq_Euler_Lagrange}
\end{align}
for $\rho(\bs{r}) = \rho_{\text{eq}}(\bs{r})$.
Equation~\eqref{Eq_Euler_Lagrange} can be solved self-consistently using (numerical) fixed-point iteration\cite{Hughes2014} at given temperature, interaction parameters, and given average density~$\bar{\rho} = \langle N \rangle / (L_x L_y)$ (where $\langle N \rangle$ is the average particle number related to the chemical potential~$\mu$).
We use periodic boundary conditions in both directions.
Some technical details are summarized in Appendix~\ref{SEC:Technical details of the DFT calculations}.

\section{Numerical results\label{SEC:Free_minimization_within_DFT}}

\subsection{Phase diagrams  \label{SUBSEC:Phase_diagrams}}
In this section, we present numerical results from free minimization of Eq.~\eqref{Eq_grand_potential_functional}.
We start by demonstrating that ultra-soft particles interacting via the GEM-4 potential indeed undergo a LIF transition on both, the cosine substrate, and the Gaussian substrate.
To this end, we perform free minimizations of $\Omega[\rho]$ for various average densities~$\bar{\rho}$ and various parameters of the external potential, $V_0$ and $R_{g}$, respectively [see Eqs.~\eqref{Eq_cosine_substrate} and \eqref{Eq_Gaussian_substrate}].
For the present substrate potentials, which vary along the $x$-direction, the simplest phase is the modulated liquid (ML) phase, which varies only along $x$ and thus can be identified by the conditions $\partial_x \rho(x,y) \neq 0$, $\partial_y \rho(x,y) = 0$.
The onset of LIF results in $\partial_y  \rho(x,y) \neq 0$.
The phase arising after LIF is a so-called locked floating solid~\cite{Bechinger2007}: It is "locked" along the $x$-direction by the 1D substrate, but can freely slide along the $y$-direction.
We will use the above criteria for $\rho(x,y)$ to categorize the density profiles obtained through free minimization. 
Throughout the calculations, we choose the substrate periodicity~$L_s$ such that every potential minimum contains lattice sites of the solid phase after freezing.

\begin{figure*}
    \centering
	
	\includegraphics[width=0.95\textwidth]{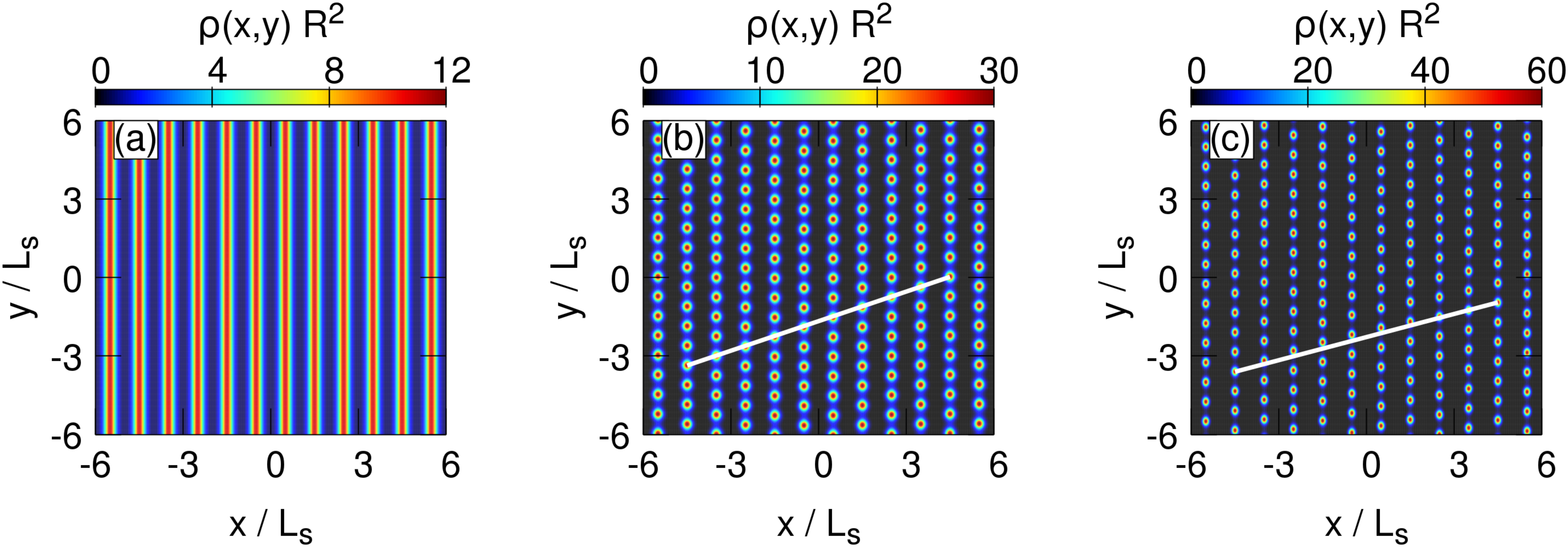}	
		
    \caption{Representative density profiles $\rho(x,y)$ before and after the onset of LIF on the cosine substrate.
			Parts (a)-(c) show results for different values of $V_0$.
			The transition occurs at $\beta V_0 = 5.4$.
			(a)~Modulated liquid phase ($\beta V_0 = 5.2$),
			(b)~locked floating solid phase ($\beta V_0 = 7$).
			The white straight line reflects the (orientational) ordering of particles between adjacent minima.
			(c)~ Results deep in the solid phase where the orientational ordering is partially lost, as indicated by the deviations of particle positions from the straight line.
			In all parts, the average density is $\bar{\rho} \, R^2= 4$, and the substrate periodicity is $L_s / R = 1.8$.
			}
    \label{Fig_classical_LIF_cosine_substrate}
\end{figure*}

In Fig.~\ref{Fig_classical_LIF_cosine_substrate}, we illustrate the LIF transition on the cosine substrate at fixed $\bar{\rho} \, R^2= 4$ by showing the density distribution before and after the transition.
At small values of $V_0$ [see Fig.~\ref{Fig_classical_LIF_cosine_substrate}(a)], the system displays the ML phase.
Upon increase of $V_0$, it freezes into the locked floating solid [Fig.~\ref{Fig_classical_LIF_cosine_substrate}(b)] thus demonstrating the occurrence of LIF.
The figure also shows that the locked floating solid is characterized by orientational order, that is, a triangular arrangement between particles in adjacent minima.
For even larger~$V_0$, however, lattice sites in adjacent minima gradually lose their orientational ordering.
This is indicated by the fact that lattice sites in different minima deviate more and more from a straight line [see white line in Fig.~\ref{Fig_classical_LIF_cosine_substrate}(b) and (c)].
At present it is unclear whether the gradual loss of orientational ordering is just an artefact of the otherwise highly accurate (at high density) mean-field approximation.
The gradual loss might also be a precursor of re-entrant melting.
However, we never observed a true re-entrant melting for the GEM-4 particles even though we performed an extensive search in the parameter space~($\bar{\rho}$, $V_0$, $L_s$).
This finding is interesting also in the broader context of the freezing of ultra-soft particles:
According to a criterion proposed by Likos \textit{et al.}\cite{Likos2001PhysRevE}, \emph{bulk systems} of ultra-soft particles first freeze and then show re-entrant melting for larger densities (below an upper freezing temperature), if the particle interactions belong to the so-called $Q^+$-class (with positive definite Fourier transform of the pair interaction).
However, if the pair interaction belongs to the $Q^{\pm}$-class (which is characterized by both, positive and negative parts in the Fourier transform, as does the GEM-4 interaction), the system freezes at sufficiently high density, but does not show re-entrant melting. 
It might be very interesting to investigate if this criterion persists for particles on patterned substrates.
However, this is beyond the scope of the present paper.

\begin{figure*}
   \centering
	\includegraphics[width=0.95\textwidth]{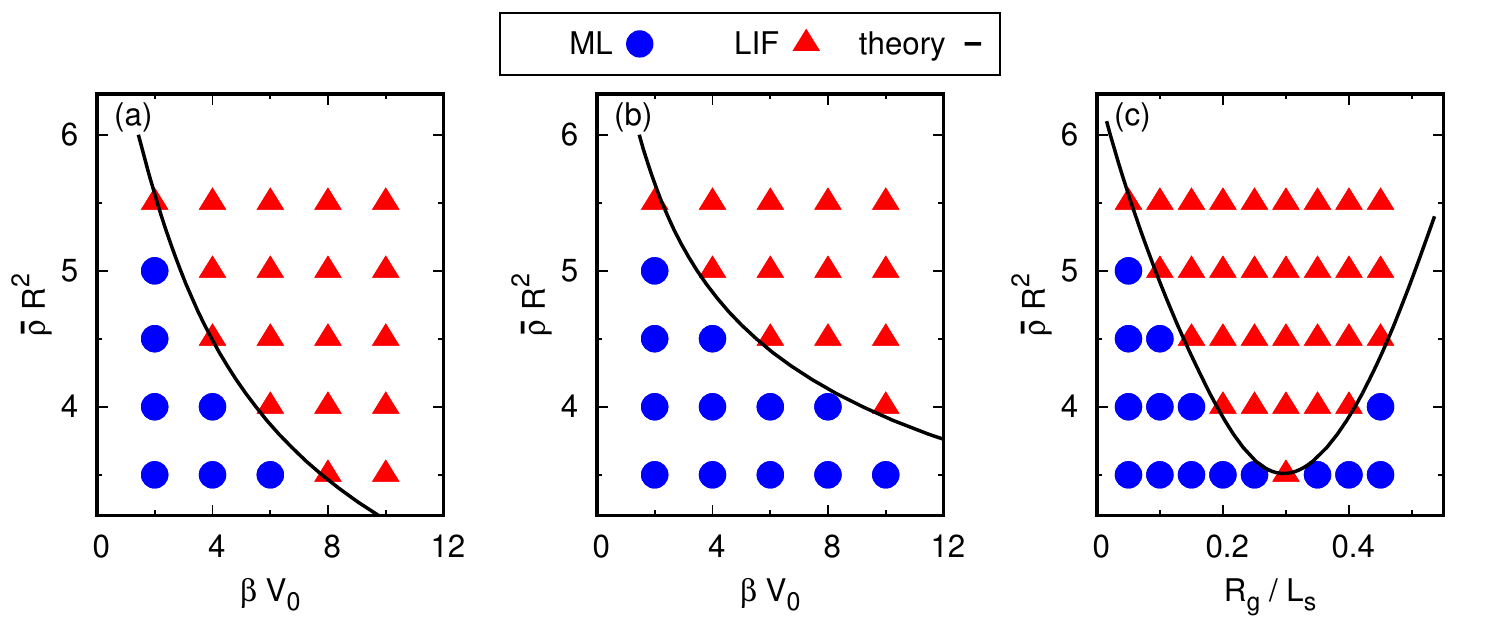}
	
	\caption{Phase diagram obtained for various average densities~$\bar{\rho}$ on 
	(a) the cosine substrate for varying potential amplitude~$V_0$, 
	(b) the Gaussian substrate for varying $V_0$ and fixed range $R_{g}$ ($R_{g}/L_s = 0.2$),  and 
	(c) the Gaussian substrate for varying $R_g$ and fixed $V_0$ ($\beta V_0 = 10$).
	The symbols correspond to results from free minimization of the density functional, while the black solid line describes the prediction from our theory (see Section~\ref{SEC:Explicit_calculations_for_our_model_system}). The substrate periodicity is $L_s / R= 1.8$.}
    
    \label{Fig_CompareParameterScanWithPrediction_L_s=1.8}
\end{figure*}

\begin{figure*}
    \centering	
	\includegraphics[width=0.95\textwidth]{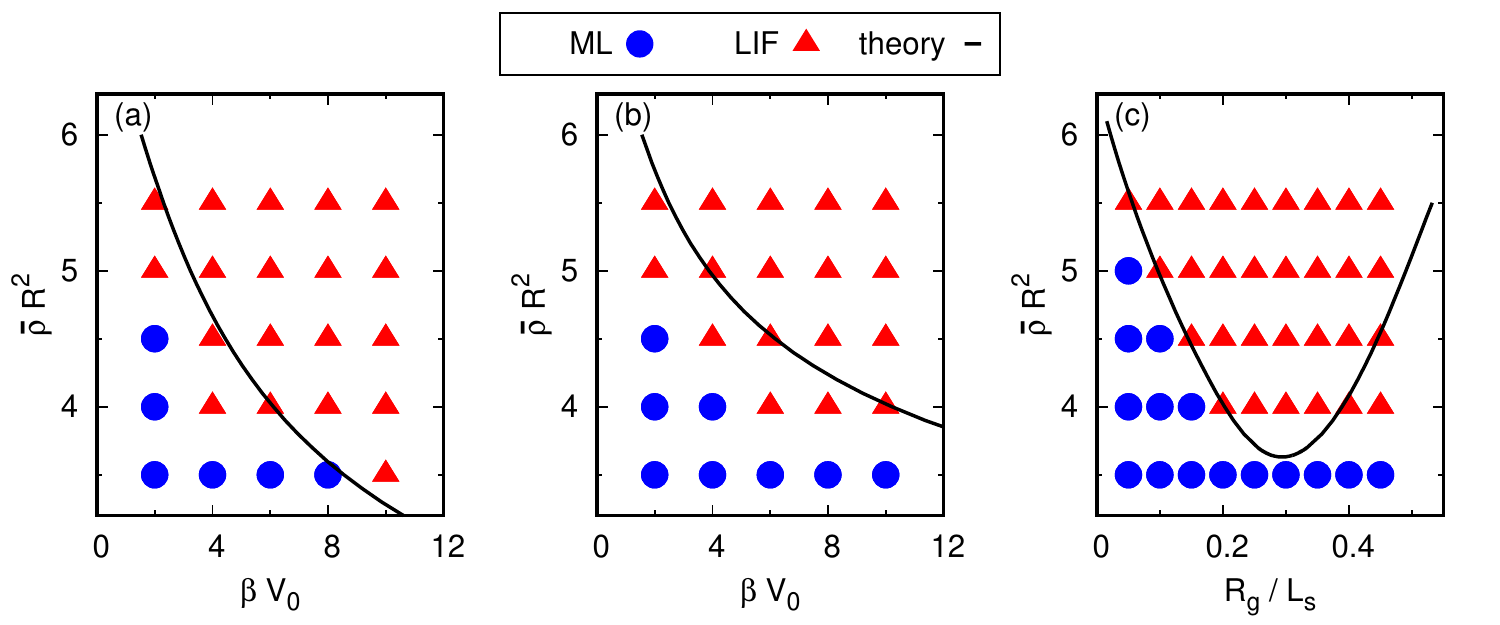}
    \caption{Same as Fig.~\ref{Fig_CompareParameterScanWithPrediction_L_s=1.8}, but for substrate periodicity $L_s / R = 1.6$.}
    \label{Fig_CompareParameterScanWithPrediction_L_s=1.6}
\end{figure*}

\begin{figure*}
    \centering	
	\includegraphics[width=0.95\textwidth]{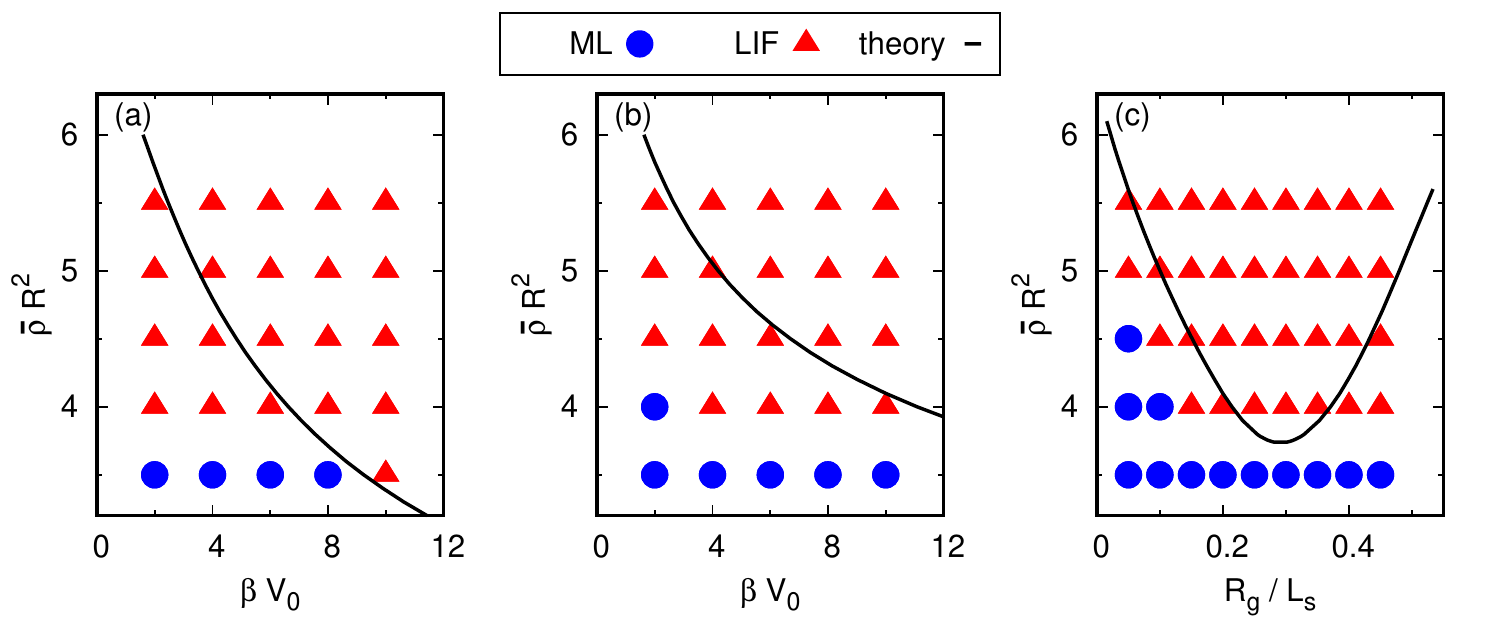}
    \caption{Same as Fig.~\ref{Fig_CompareParameterScanWithPrediction_L_s=1.8}, but for substrate periodicity $L_s / R = 1.2$.}
    \label{Fig_CompareParameterScanWithPrediction_L_s=1.2}
\end{figure*}

We now turn back to the LIF transition.
Having demonstrated that LIF occurs on the cosine substrate (see Fig.~\ref{Fig_classical_LIF_cosine_substrate}), we scanned large portions of the parameter space ($\bar{\rho}$, $V_0$) for the cosine substrate  and the Gaussian substrate (with variable $V_0$ or variable $R_g$) at periodicities $L_s / R= 1.8$, $L_s / R= 1.6$, and $L_s / R= 1.2$.
Results are presented in Fig.~\ref{Fig_CompareParameterScanWithPrediction_L_s=1.8}, Fig.~\ref{Fig_CompareParameterScanWithPrediction_L_s=1.6}, and Fig.~\ref{Fig_CompareParameterScanWithPrediction_L_s=1.2} respectively.
We note that the numerical results are based on visual inspection of the corresponding spatial configurations described by $\rho(\bs{r})$.
From these, we found no indications for a first-order transition, the changes between the two phases appeared rather smooth. (However, we did not investigate this issue systematically.)
For both, the cosine substrate and the Gaussian substrate with fixed $R_g$, we observe a LIF transition at sufficiently large values of $V_0$.
Furthermore, for the Gaussian substrate at constant $\beta V_0 = 10$, we varied the available space around the potential minima through the range $R_{g}$ of the Gaussian peak.
As shown in Fig.~\ref{Fig_CompareParameterScanWithPrediction_L_s=1.8}(c) and Fig.~\ref{Fig_CompareParameterScanWithPrediction_L_s=1.6}(c), this leads to freezing as well.
Note that for large values of $R_{g}$, the overlap of two Gaussian maxima becomes increasingly important and decreases the potential difference between maxima and minima.
Thus, the bending-up of the LIF transition curves [Fig.~\ref{Fig_CompareParameterScanWithPrediction_L_s=1.8}(c), Fig.~\ref{Fig_CompareParameterScanWithPrediction_L_s=1.6}(c), and \ref{Fig_CompareParameterScanWithPrediction_L_s=1.2}(c)] is not attributed to re-entrant melting, but due to an effectively reduced potential difference.
We also studied the influence of $L_s / R$ on LIF for periodicities  $L_s / R = 1.4$, and $1.0$.
Comparing different periodicities with regard to the onset of LIF, we find  a (slight) preference of the value $L_s / R = 1.2$. 
This is consistent with our expectation that freezing in the 1D potential occurs most likely when the locked floating solid with the lattice constant equal to the bulk lattice constant "optimally fits" into the substrate in its primary orientation\cite{Bechinger2007}.
For the present system, the (2D) bulk lattice constant is $a / R \approx 1.4$, yielding \mbox{$L_s/R = \sqrt{3}a /(2 R) \approx 1.2$} as an optimal value.

For later reference, we have also included our theoretical prediction for the onset of LIF in Figs.~\ref{Fig_CompareParameterScanWithPrediction_L_s=1.8},  \ref{Fig_CompareParameterScanWithPrediction_L_s=1.6}, and \ref{Fig_CompareParameterScanWithPrediction_L_s=1.2} (black solid line).
The theory behind the prediction is outlined in the subsequent sections, see particularly Eq.~\eqref{Eq_EOS_form_averaged_stress_within_effective_density_ansatz}.
We remark already here that the theoretical prediction does not contain any fitting parameters.
Above the black solid line, the system is in the solid phase.
We find excellent agreement with the DFT data in the case $L_s / R = 1.8$, as seen in Fig.~\ref{Fig_CompareParameterScanWithPrediction_L_s=1.8}.
At the slightly smaller substrate periodicity $L_s / R = 1.6$ (Fig.~\ref{Fig_CompareParameterScanWithPrediction_L_s=1.6}), the theoretical prediction still provides a good estimate for the onset of LIF.
This situation somewhat changes at $L_s / R = 1.2$ (Fig.~\ref{Fig_CompareParameterScanWithPrediction_L_s=1.2}) where the solid line is located deeply within the numerically obtained LIF regime.
We will provide a corresponding argument in Sec.~\ref{SEC:Explicit_calculations_for_our_model_system}.

\subsection{LIF as a density-driven transition \label{SUBSEC:LIF_as_a_density_driven_transition}}

\begin{figure}
    \centering
    \includegraphics[width=0.45\textwidth]{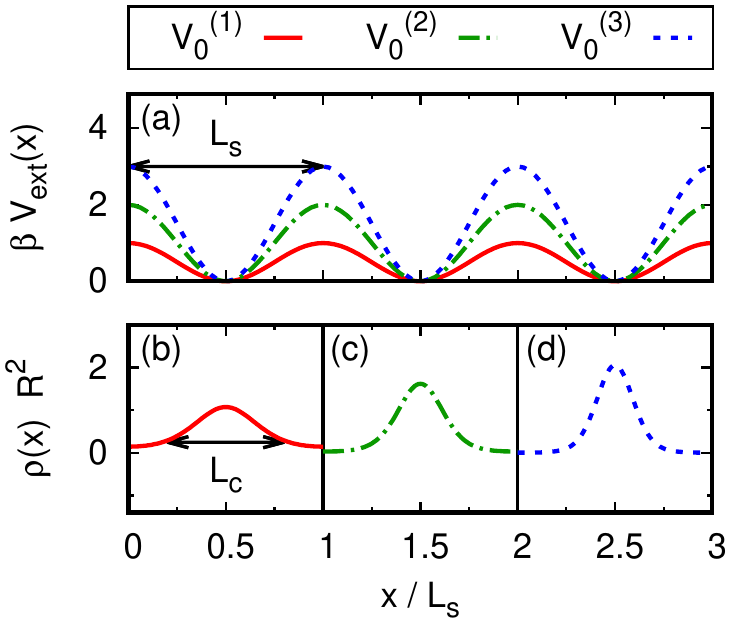}
    \caption{Illustration of the concept of the confining length~$L_c$ on the cosine substrate. 
		(a) Cosine potential for different amplitudes $V_0$, shifted such that the minima overlap. Also the substrate periodicity $L_s$ is shown. 
		Parts (b)-(d) show density profiles for increasing values of $V_0$.		
		With increasing $V_0$, the vicinity of the minima changes which results in a reduction of the confining length $L_{c}$ of the particle distribution $\rho(\bs{r})$. 
    }
	\label{Fig_cosine_for_different_V_0_shifted}
\end{figure}

In this section we aim at demonstrating that the LIF transition, which is seemingly controlled by the potential amplitude $V_0$, can be closely related to corresponding changes in the microscopic density profile $\rho(\bs{r})$.
To this end, we introduce two new parameters.
To motivate this step, we recall that, due to periodicity, the \emph{average} density in one modulation of the periodic potential is the same as the average density $\bar{\rho}$ of the (nonetheless inhomogeneous) system.
However, the latter is not a representative quantity, especially along the $x$-direction, since particles preferably occupy regions around the potential minima. 
Thus, within one modulation, a large fraction $f$ of particles are located within a "confining" length~$L_{c}$, which is smaller than the substrate periodicity $L_{s}$.
The definition of the confining length $L_c$ is illustrated in Fig.~\ref{Fig_cosine_for_different_V_0_shifted}.
For a given value of $f$, we define $L_c$ according to
\begin{align}
 \int\limits _{-\infty}^{\infty} dy  \int\limits _{x_{\text{min}} -  \frac{L_{c}}{2} }^{x_{\text{min}} +  \frac{L_{c}}{2} } dx\,  \rho(x,y)   =  f  \int\limits_{-\infty}^{\infty} dy  \int\limits_{x_{\text{min}} -  \frac{L_{s}}{2} }^{x_{\text{min}} +  \frac{L_{s}}{2} } dx\, \rho(x,y).
\label{Def_L_c}
\end{align}
Clearly, the choice of $f$ requires some consideration, which will be given later.
Here, we only note that $f$ should be less than one, since that would correspond to $L_c=L_s$.

Besides $L_c$, the second new quantity characterizing the density distribution $\rho(\bs{r})$ in the vicinity of the minima is given by
\begin{align}
	\bar{\rho}_{\text{eff}} =  \frac{1}{A_c} \int\limits _{-\infty}^{\infty} dy  \int\limits _{x_{\text{min}} -  \frac{L_{c}}{2} }^{x_{\text{min}} +  \frac{L_{c}}{2} } \hspace{-1.5em} dx  \, \rho(x,y),
	\label{Def_rho_eff}
\end{align}
corresponding to the effective average density within the region enclosed by $L_c$ (with area $A_c = L_c L_y$ with \mbox{$L_y \to \infty$}).

\begin{figure*}
    \centering
   \includegraphics[width=\textwidth]{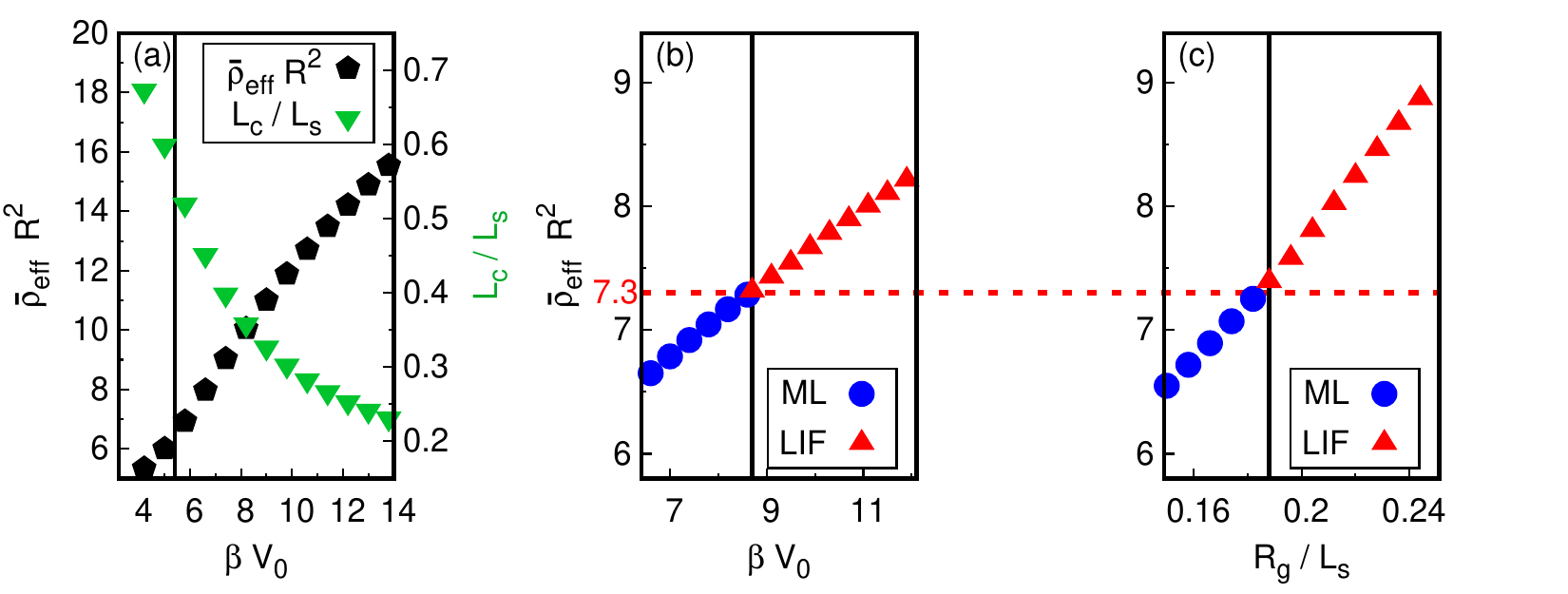}
		
    \caption{(a) The parameter $\bar{\rho}_{\text{eff}}$ and $L_{c}$ as functions of the potential amplitude~$V_0$ on the cosine substrate at average density $\bar{\rho} \, R^2 = 4$ (see also Fig.~\ref{Fig_classical_LIF_cosine_substrate}).
			For small values of $V_0$, the system is in the modulated liquid phase, whereas LIF arises for larger values. 
			The black vertical line indicates the onset of LIF in the DFT calculations ($\beta V_0= 5.4$).	
			Note that the two curves belong to different axes and their crossing has no physical meaning.
			(b) and (c) show the variation of the effective average density $\bar{\rho}_{\text{eff}}$ upon change of the two parameters of the Gaussian substrate potential at $\bar{\rho} \, R^2 = 4$.
    	In~(b), the substrate amplitude~$V_0$ is varied at fixed \mbox{$R_{g}/L_{s} = 0.2$}, whereas in (c) the available space is varied via $R_{g}$ at fixed $\beta V_0 = 10$.	
	The black vertical line represents the onset of LIF in the DFT calculations at $\beta V_0= 8.7$ and  \mbox{$R_{g}/L_{s}=  0.188$}, respectively. 
	LIF sets in at approximately the same threshold effective average density \mbox{$\bar{\rho}_{\text{eff}} R^2 \approx 7.3$} (see red dashed line).
    }
    \label{Fig_L_confined_and_rho_eff_vs_V_0_and_R_g}
\end{figure*}

We now argue that the parameters $L_c$ and $\bar{\rho}_{\text{eff}}$ can indeed be considered as new control parameters for the LIF transition.
To this end, we revisit our study of LIF on the cosine substrate at average density $\bar{\rho} \,R^2 = 4$ (see Fig.~\ref{Fig_classical_LIF_cosine_substrate}).
For fixed (yet arbitrary) $f = 0.9$, we can extract $L_{c}$ numerically and consequently determine $\bar{\rho}_{\text{eff}}$ (for each value of $V_0$) from the obtained density distributions~$\rho(\bs{r})$.
The results are shown in Fig.~\ref{Fig_L_confined_and_rho_eff_vs_V_0_and_R_g}(a).
It is seen that an increase of the potential amplitude $V_0$ results in a reduction of the confining length $L_{c}$, as already indicated by Fig.~\ref{Fig_cosine_for_different_V_0_shifted}.
This reflects the redistribution of particles from unfavourable positions, i.e. the potential maxima, to the vicinity of the minima.
This, in turn, leads to an increase of the effective average density $\bar{\rho}_{\text{eff}}$ within the minima.
The increase of the latter finally generates spontaneous symmetry breaking from the ML phase into the phase with $\partial_y \rho(x,y) \neq 0$.
We have repeated this kind of calculation for different average densities $\bar{\rho} \, R^2$ (data not shown).
It turns out that the density $\bar{\rho}_{\text{eff,c}}$, at which LIF occurs, does not vary substantially.

In summary, these results already suggest that the LIF phase transition might be considered as a density-driven phase transition, with a threshold value of $\bar{\rho}_{\text{eff}}$ that, upon exceeding, leads to the LIF transition. 
The next step in our argumentation is a consistency check:
If the LIF transition can indeed be related to an increase of $\bar{\rho}_{\text{eff}}$, one would expect $\bar{\rho}_{\text{eff,c}}$ to be independent of \emph{how} the density increased.
To show that this is indeed the case, we consider the Gaussian substrate and explore two independent variants to increase $\bar{\rho}_{\text{eff}}$.
Results are shown in Fig.~\ref{Fig_L_confined_and_rho_eff_vs_V_0_and_R_g}(b) and ~\ref{Fig_L_confined_and_rho_eff_vs_V_0_and_R_g}(c).
First, in Fig.~\ref{Fig_L_confined_and_rho_eff_vs_V_0_and_R_g}(b), we increase the potential amplitude~$V_0$ as in the conventional LIF scenario.
We observe that the effective average density $\bar{\rho}_{\text{eff}}$ gradually increases and for $\beta V_0 \geq 8.7$, we observe light-induced freezing.
Second, in Fig.~\ref{Fig_L_confined_and_rho_eff_vs_V_0_and_R_g}(c), at constant $\beta V_0=10$, we decrease the available space by increasing the range $R_{g}$ of the Gaussian peak.
This enhances $\bar{\rho}_{\text{eff}}$ by purely geometric means. 
Note that although the substrate potential amplitude $\beta V_0=10$ is larger in the second variant, LIF only occurs at \mbox{$R_{g}/L_{s} \geq 0.188$}.
However, the threshold density  $\bar{\rho}_{\text{eff,c}}$ at the transition has essentially the same value ($\bar{\rho}_{\text{eff}} R^2 \approx 7.3$) as in the first case [cf. red line in Figs.~\ref{Fig_L_confined_and_rho_eff_vs_V_0_and_R_g}(b) and (c)].
This strongly supports the role of $\bar{\rho}_{\text{eff}}$ as a control parameter.

One somewhat weak point of our analysis so far concerns the choice of the parameter $f$.
Until now we have (arbitrarily) set $f=0.9$.
Whereas the basic mechanism of a density-driven LIF transition remains true for any choice of $f$, the actual numerical values of $L_c$ and $\bar{\rho}_{\text{eff}}$ will clearly depend on $f$.
In the following section, we propose one possible way to circumvent the problem of first obtaining the density profile $\rho(\bs{r})$ and then obtaining $L_c$ and $\bar{\rho}_{\text{eff}}$ based on a specific value of $f$.

\section{Towards a theoretical description\label{SEC:Towards_a_theoretical_description}}

\subsection{Density functional relations and density parametrization}
Our goal is to establish a relation between $\makeEffectiveAverageDensitySymbol$, $L_c$, and $V_0$ based on density functional arguments.
Our starting point is the balance equation of hydrostatics~\cite{Evans1979} 
\begin{align}
 \nabla \cdot \bs{\sigma}(\bs{r}) = \rho(\bs{r}) \nabla V_{\text{ext}}(\bs{r}),
 \label{Eq_Cauchy_momentum_equation_stationary}
\end{align}
where $\bs{\sigma}$ denotes the (second-order) stress tensor, which is the negative of the usual pressure tensor\cite{Henderson1992}.
Equation~\eqref{Eq_Cauchy_momentum_equation_stationary} expresses the fact that the stress inside the system is balanced by the force stemming from the external potential.
Using the equilibrium condition~\eqref{Eq_Euler_Lagrange}, the right side of \eqref{Eq_Cauchy_momentum_equation_stationary} can be expressed via the functional derivative of the intrinsic Helmholtz free energy,
\begin{equation}
	\nabla \cdot \bs{\sigma}(\bs{r}) = - \rho(\bs{r}) \nabla  \left( \frac{\delta F[\rho]}{\delta \rho(\bs{r})}\right),
	\label{Eq_Divergence_of_stress_tensor_from_DFT}
\end{equation}
where \mbox{$F= F_{\text{id}} + F_{\text{exc}}$}  (see Section \ref{SUBSEC:Model_and_density_functional_theory_subsection_DFT}).
Evaluating the functional derivative (before making any approximations for $F_{\text{exc}}$), the right side of Eq.~\eqref{Eq_Divergence_of_stress_tensor_from_DFT} involves the gradient of the one-point direct correlation function, $c^{(1)}(\bs{r}) = -\beta \delta F_{\text{exc}}[\rho]/ \delta \rho(\bs{r})$.
We also note that this rewriting is equivalent to expressing \mbox{$\rho \nabla V_{\text{ext}}$} in Eq.~\eqref{Eq_Cauchy_momentum_equation_stationary} via the Lovett-Mou-Buff-Wertheim (LMBW) equation~\cite{Lovett1976, Wertheim1976}, which yields
\begin{align}
\label{Eq_Divergence_of_stress_tensor_from_LMBW}
 \nabla \cdot \bs{\sigma}(\bs{r}) &= - \beta^{-1} \nabla \rho(\bs{r}) \notag\\
									&\quad+ \rho(\bs{r}) \beta^{-1}  \int d \bs{r}' c^{(2)}(\bs{r}, \bs{r}') \nabla' \rho(\bs{r}').
\end{align}
In Eq.~\eqref{Eq_Divergence_of_stress_tensor_from_LMBW}, $c^{(2)}(\bs{r}, \bs{r}')$ denotes the two-particle direct correlation function related to the density $\rho(\bs{r})$.
All of these equations are exact, and given the true equilibrium density and the true correlations, Eqs.~\eqref{Eq_Cauchy_momentum_equation_stationary}, \eqref{Eq_Divergence_of_stress_tensor_from_DFT} and \eqref{Eq_Divergence_of_stress_tensor_from_LMBW} yield consistent results.
For reasons outlined below, we here consider an integrated form of Eq.~\eqref{Eq_Cauchy_momentum_equation_stationary}, that is,
\begin{align}
 \label{Eq_averaged_stress}
 \frac{1}{\mathcal{A}} \int d\mathcal{A}\, \text{sign}(x)\, &\bs{e}_x \, \nabla \cdot\bs{\sigma} \notag\\
 &= \\ 
 \frac{1}{\mathcal{A}} \int d\mathcal{A}\, \text{sign}(x)\, &\bs{e}_x \rho(\bs{r}) \nabla V_{\text{ext}}(x) ,\notag
\end{align}
where $\bs{e}_x$ denotes the unit vector in $x$-direction and $\text{sign}(\cdots)$ denotes the sign function.
The integration is performed over an area $\mathcal{A}$ in the $x$-$y$ plane, and the resulting integrals are then divided by that area (details outlined below).
Specifically, we focus on a region $\mathcal{A} = [-\frac{L_x}{2}, \frac{L_x}{2}] \times [-\frac{L_y}{2}, \frac{L_y}{2}]$ centered around the minimum of the substrate potential $V_{\text{ext}}(x)$, say $x=0$ (for notational convenience).
We note that the quantities involved in Eq.~\eqref{Eq_Cauchy_momentum_equation_stationary} are anti-symmetric with respect to the location of the minimum, such that a direct average would result to zero.
We thus multiply both sides of Eq.~\eqref{Eq_Cauchy_momentum_equation_stationary} by $\text{sign}(x)$.
We further multiply with $\bs{e}_x$, since we are interested in the $x$-component of the force.
Equation~\eqref{Eq_averaged_stress} is still exact.
This will be our starting point for approximations.

Our strategy towards an theoretical description of the onset of LIF is now as follows:
Starting from Eq.~\eqref{Eq_averaged_stress}, we evaluate both sides on the basis of an \textit{ansatz} for the density profile, which involves the parameters 
$\makeEffectiveAverageDensitySymbol$ and $L_c$ introduced in Section \ref{SUBSEC:LIF_as_a_density_driven_transition}.
The resulting \textit{approximate} equation then gives a relation between the density parameters, the external potential, and the stress [via the left side of Eq.~\eqref{Eq_averaged_stress}]  or, respectively, the correlations of the system.
To use this relation in the context of LIF, we compare $\makeEffectiveAverageDensitySymbol$ to the density where the \textit{bulk} system becomes unstable against freezing. This finally yields a prediction for the onset of LIF, that is, 2D freezing in the presence of a 1D substrate potential.

To apply this strategy, we work with the following (strongly simplified) ansatz for the effective density profile $\rho(x)$ in the vicinity of the substrate minima (say, $x=0$),
\begin{align}
\rho(x) 	= \bar{\rho}_{\text{eff}} \, \text{rect}\left(\frac{x}{L_c}\right) &= \begin{cases}
        	                                                                 \bar{\rho}_{\text{eff}} & \text{if }  |x| \leq \frac{L_c}{2} \\
        	                                                                 0			, &	\text{else}.
        	                                                                \end{cases} \label{Eq_effective_density_ansatz}
\end{align}
The ansatz contains the width $L_c$ and the height $\bar{\rho}_{\text{eff}}$ as parameters, which are linked by the condition of conservation of particles, that is,
\begin{align}
	\label{Eq_redistribution_of_particles}
	\bar{\rho}_{\text{eff}} = \bar{\rho} \cdot \frac{L_s}{L_c}.
\end{align}
The ansatz is periodically repeated with the substrate periodicity $L_s$ (see Appendix~\ref{SEC:Details_of_the_effective_density_profile_parametrization} for technical details).
Clearly, the (rectangular) ansatz~\eqref{Eq_effective_density_ansatz} for the density profile does not fulfill the exact balance equation~\eqref{Eq_Cauchy_momentum_equation_stationary} and its variants \eqref{Eq_Divergence_of_stress_tensor_from_DFT}, \eqref{Eq_Divergence_of_stress_tensor_from_LMBW}.
However, here we are working with the integrated form [Eq.~\eqref{Eq_averaged_stress}], where the impact of the approximation is less obvious.
On the one hand, one would still expect inconsistencies when expressing $\divergence \bs{\sigma}$ in different ways, just because of the approximate nature of Eq.~\eqref{Eq_effective_density_ansatz} (similar to the related problem of thermodynamic inconsistency when evaluating, e.g., the pressure by different routes\cite{HansenMcDonald}).
On the other hand, one could interpret Eq.~\eqref{Eq_averaged_stress} in the sense of the mean value theorem for integrals\cite{Larson2005calculus}. To this end we note that both integrals appearing in Eq.~\eqref{Eq_averaged_stress} are functionals of the profile $\rho(\bs{r})$.
The mean value theorem then states that there is a mean profile (in the space of possible profiles) such that the value of the integrals evaluated with this mean profile (which we here choose to be rectangular) is representative of this integral.
From this perspective, our ansatz may appear somewhat less unplausible.

We now consider in detail the two sides of Eq.~\eqref{Eq_averaged_stress} in combination with the parametrized density \eqref{Eq_effective_density_ansatz}.
The right side can be evaluated directly, yielding
	\begin{align}
		I_V 	& = \frac{1}{L_x L_y} \int\limits_{-L_c/2}^{L_c/2} \!\!\!\!\!dx\, \int\limits_{-L_y/2}^{L_y/2} \!\!\!\!\!dy\,\, \text{sign}(x)\, \bs{e}_x \, \makeEffectiveAverageDensitySymbol \nabla V_{\text{ext}}(x) \notag\\
			&=\frac{1}{L_x}  2\, \bar{\rho}_{\text{eff}}  V_{\text{ext}}\left(\frac{L_c}{2} \right),
	\label{Eq_averaged_external_stress_within_effective_density_ansatz}
	\end{align}
where we have assumed a symmetric and appropriately shifted external potential such that $V_{\text{ext}}(x) = V_{\text{ext}}(-x)$, $V_{\text{ext}}(0) = 0$. 

The evaluation of the left side of Eq.~\eqref{Eq_averaged_stress},
	\begin{align}
			I_{\bs{\sigma}} \equiv \frac{1}{\mathcal{A}} \int d\mathcal{A}\, \text{sign}(x)\, &\bs{e}_x \, \nabla \cdot\bs{\sigma}
			\label{Eq_Integration_of_div_stress_tensor}
	\end{align}
is less straightforward due to the more involved dependency of the integrand on the density profile [see, e.g., Eq.~\eqref{Eq_Divergence_of_stress_tensor_from_LMBW}].

\section{Explicit calculations for our model system\label{SEC:Explicit_calculations_for_our_model_system}}

In this section, we discuss three different variants to evaluate the integral $I_{\bs{\sigma}}$ containing the stress tensor, Eq.~\eqref{Eq_Integration_of_div_stress_tensor}.
Combining then $I_{\bs{\sigma}}$ with $I_{V}$ [according to Eq.~\eqref{Eq_averaged_stress}], this yields different relations between the density parameters and the external potential.
We then use these relations for our prediction of LIF, as outlined below.
Since it is not \emph{a priori} clear which variant produces the best prediction, we show the LIF prediction of all variants and compare them with our previous results from free minimization of the density functional (see Figs.~\ref{Fig_CompareParameterScanWithPrediction_L_s=1.8}, \ref{Fig_CompareParameterScanWithPrediction_L_s=1.6}, and \ref{Fig_CompareParameterScanWithPrediction_L_s=1.2}).
Consistent with these calculations, we use the mean-field approximation in the different variants of 
evaluating $I_{\bs{\sigma}}$ as well.
The corresponding results are shown in Fig.~\ref{Fig_parameterScan_and_different_variants_of_prediction}.

It is instructive to first present the simplest and most straightforward variant where $\divergence \bs{\sigma}$ is expressed via the right side of Eq.~\eqref{Eq_Divergence_of_stress_tensor_from_DFT}
(even though we will later see that this strategy does not perform very well.)

\subsection{Evaluation via Eq.~\eqref{Eq_Divergence_of_stress_tensor_from_DFT}\label{SEC:Prediction_variant_A}}
In this variant, we express the derivative $\delta F[\rho]/\delta \rho(\bs{r})$ appearing on the right side of Eq.~\eqref{Eq_Divergence_of_stress_tensor_from_DFT} using the mean-field-approximation for the excess contribution [see Eq.~\eqref{Eq_excess_free_energy_functional}].
We then substitute the effective density profile ansatz [see Eq.~\eqref{Eq_effective_density_ansatz}] as an approximation for the density profile~$\rho(\bs{r})$.
Multiplying by $\text{sign}(x) \bs{e}_x$ and performing the integral yields
\begin{align}
	\tilde{I}_{\bs{\sigma}}  &= 2 \beta^{-1} \makeEffectiveAverageDensitySymbol  - \makeEffectiveAverageDensitySymbol^2 \sum_{j = -N_r}^{N_r} \notag\\
	& \left[ \int \limits_{-L_c/2}^{L_c/2} \!\!\!\!\!\!dx\,  \int \limits_{j L_s - L_c/2}^{j L_s + L_c/2} \!\!\!\!\!\!\!\!\!\!dx'\,\,\,\, \int \limits_{-\infty}^{\infty} \!\!\!dy'\, \text{sign}(x) \partial_x V(x-x', y') \right]
	\label{Eq_Integration_of_div_stress_tensor_variant_i}
\end{align}
where $\tilde{I}_{\bs{\sigma}}  = I_{\bs{\sigma}} L_x$.  
Setting $\tilde{I}_{\bs{\sigma}}$ equal to $\tilde{I}_{V} = I_{V} L_x$ [see Eq.~\eqref{Eq_averaged_external_stress_within_effective_density_ansatz}], we obtain a relation between the parameters ($L_c$, $\makeEffectiveAverageDensitySymbol$) and the parameters of the external potential.

We now turn to the second main step of our prediction for the onset of LIF.
As shown by our free DFT minimizations described in Sec.~\ref{SUBSEC:LIF_as_a_density_driven_transition}, the effective average density~$\bar{\rho}_{\text{eff}}$ can be considered as a driving parameter for the LIF phase transition (see Fig.~\ref{Fig_L_confined_and_rho_eff_vs_V_0_and_R_g}).
However, we do not have \textit{a priori} knowledge about the critical value $\bar{\rho}_{\text{eff,c}}$ above which LIF occurs.
As a first rough estimate, we identify $\bar{\rho}_{\text{eff,c}}$ with the density where the corresponding bulk system (in the absence of an external potential) becomes unstable.
According to Ref.~\onlinecite{Archer2014}, this occurs at $\bar{\rho}_{\text{LSA}} \, R^2= 6.38$ for our chosen parameters.
We thus set the critical value \mbox{$\bar{\rho}_{\text{eff,c}} \, R^2 = 6.38$}.
We deliberately chose the instability threshold rather than the coexistence density (which is somewhat smaller) for $\bar{\rho}_{\text{eff,c}}$. 
In this way we ensure that an effective fluid at the same density is surely unstable.
Thus the prediction should be seen as a sufficient criterion.

The equation $I_{\bs{\sigma}} = I_{V}$ with $\makeEffectiveAverageDensitySymbol = \bar{\rho}_{\text{eff,c}}$ is then solved for the substrate potential amplitude~$V_0$ in the conventional LIF transition (where $V_0$ is varied), whereas for the alternative variant of the Gaussian substrate, it is solved for the range of the Gaussian maxima $R_{g}$.
The former case can be treated explicitly (see Appendix \ref{Onset_of_LIF_for_varying_V_0}), whereas the latter has to be solved numerically.
Comparing the resulting prediction [which we call variant (A)] for the onset of LIF to the DFT data (see Fig.~\ref{Fig_parameterScan_and_different_variants_of_prediction}), however, we find that it gives rather poor results.

\subsection{Evaluation via stress tensor}

We now consider a route which focuses more explicitly on the stress tensor, $\bs{\sigma}$, appearing on the left side of Eq.~\eqref{Eq_averaged_stress}.
Following Ref.~\onlinecite{Long1961MechanicsOfSolidsAndFluids}, we decompose $\bs{\sigma}$ as
\begin{align} 
 \bs{\sigma} &= -p\, \bs{1} + \bs{\tau}
 \label{Eq_Cauchy_stress_tensor_decomposition} 
\end{align}
where the first contribution involves the (hydrostatic) pressure~$p$, a scalar isotropic quantity, $\bs{1}$ is the unit tensor, and the second (tensorial) contribution $\bs{\tau}$ represents all deviations thereof.
This tensor is called the deviatoric stress tensor\cite{Long1961MechanicsOfSolidsAndFluids}.
In the spirit of Ref.~\onlinecite{Evans1979}, we consider $p$ as a \emph{local} hydrostatic pressure, which is space-dependent and can be identified with the negative of the grand potential density, $\omega(\bs{r})$. 
This yields
\begin{align}
 p(\bs{r}) &= -\omega(\bs{r}) \notag \\
		&= -\big[ f(\bs{r}, [\rho]) + \rho(\bs{r}) V_{\text{ext}}(\bs{r})  - \mu \rho(\bs{r}) \big] \notag\\
		&= 	\big[\mu  -  V_{\text{ext}}(\bs{r}) \big]\rho(\bs{r}) - f(\bs{r}, [\rho])
		\label{Eq_local_hydrostatic_pressure}
\end{align}
where $f(\bs{r}, [\rho])$ is the intrinsic Helmholtz free energy density related to the intrinsic Helmholtz free energy, \mbox{$F= \int d \bs{r}\, f(\bs{r})$}, introduced below Eq.~\eqref{Eq_grand_potential_functional}.
In equilibrium, one has the well-known DFT relation
\begin{equation}
	\mu = V_\text{ext}(\bs{r}) + \frac{\delta F[\rho]}{\delta \rho(\bs{r})},
	\label{Eq_OmegaFunctionalDerivative}
\end{equation}
where ${\delta F[\rho]}/{\delta \rho(\bs{r})} \equiv \mu(\bs{r})$ can be regarded as the intrinsic chemical potential~\cite{Evans1979, HansenMcDonald, Evans1992}.
Equation~\eqref{Eq_OmegaFunctionalDerivative} is equivalent to the Euler-Lagrange equation~\eqref{Eq_Euler_Lagrange}.
With Eq.~\eqref{Eq_OmegaFunctionalDerivative}, the local hydrostatic pressure [see Eq.~\eqref{Eq_local_hydrostatic_pressure}] becomes
\begin{align}
	p(\bs{r})	&= \left(\frac{\delta F[\rho]}{\delta \rho(\bs{r})}\right)  \rho(\bs{r}) -  f(\bs{r}, [\rho]).
	\label{Eq_local_hydrostatic_pressure_given_by_intrinsic_quantities}
\end{align}
We now turn back to Eq.~\eqref{Eq_Cauchy_stress_tensor_decomposition}, from which it directly follows that
\begin{align} 
  \nabla \cdot \bs{\sigma} &= -\nabla p + \nabla \cdot  \bs{\tau}.
 \label{Eq_Cauchy_stress_tensor_decomposition_take_divergence} 
\end{align}
Using Eqs.~\eqref{Eq_Cauchy_momentum_equation_stationary} and \eqref{Eq_local_hydrostatic_pressure_given_by_intrinsic_quantities}, we obtain for the divergence of the deviatoric stress
\begin{align}
 \nabla \cdot \bs{\tau} 	&= \divergence \bs{\sigma} + \nabla p \notag \\
					&= \rho(\bs{r}) \nabla V_{\text{ext}}(\bs{r}) + \nabla p 	\notag\\
					&=  \left(\frac{\delta F[\rho]}{\delta \rho(\bs{r})}\right)  \nabla \rho(\bs{r}) - \nabla f(\bs{r}, [\rho]).
 \label{Eq_deviatoric_stress_tensor_given_by_intrinsic_quantities}
\end{align}
For a homogeneous bulk system in equilibrium, $p(\bs{r})$ corresponds to the bulk pressure and one would expect \mbox{$\divergence \bs{\tau}$} to vanish. 
This is indeed the case, since all gradients in Eq.~\eqref{Eq_deviatoric_stress_tensor_given_by_intrinsic_quantities} result to zero.

So far, the expressions for the contribution to $\divergence \bs{\sigma}$ are completely general.
We now specialize to the present (i.e. ultra-soft) system, which we treat in the MF approximation, such that 
\begin{align}
f(\bs{r}, [\rho]) =&\beta^{-1} \rho(\bs{r}) \left[ \ln(\Lambda^2 \rho(\bs{r}))-1\right] \notag\\&+ \frac{1}{2} \int d\bs{r}' \big[ \rho(\bs{r})  V(\bs{r}-\bs{r}')  \rho(\bs{r}') \big].
\end{align}
With this, we find from Eqs.~\eqref{Eq_local_hydrostatic_pressure_given_by_intrinsic_quantities} and \eqref{Eq_deviatoric_stress_tensor_given_by_intrinsic_quantities}
\begin{alignat}{3}
p(\bs{r}) &=  \beta^{-1} \rho(\bs{r})  &+\frac{1}{2} \int d\bs{r}'\, \rho(\bs{r}) \, V(\bs{r} - \bs{r}') \, \rho(\bs{r}')  \label{Eq_local_hydrostatic_pressure_mean_field_approximation} 
\intertext{and}
\nabla \cdot \bs{\tau}(\bs{r}) &= 		&\frac{1}{2} \int d\bs{r}'\, \nabla \rho(\bs{r})\,   V(\bs{r} - \bs{r}')\,   \rho(\bs{r}') \notag \\
							&&-\frac{1}{2} \int d\bs{r}'\,  \rho(\bs{r})\,  \nabla V(\bs{r} - \bs{r}')\,   \rho(\bs{r}'). \label{Eq_divergence_of_deviatoric_stress_tensor_mean_field_approximation} 
\end{alignat}

We now come back to Eq.~\eqref{Eq_Integration_of_div_stress_tensor}.
Decomposing the stress tensor as discussed above, we have
	\begin{align}
		I_{\bs{\sigma}} &= I_p + I_{\bs{\tau}} \\
	\intertext{where}
		I_p &\equiv \frac{1}{\mathcal{A}} \int d\mathcal{A}\, \text{sign}(x)\, \bs{e}_x \, (- \nabla p) \\
		\intertext{and}
		I_{\bs{\tau}} &\equiv  \frac{1}{\mathcal{A}} \int d\mathcal{A}\, \text{sign}(x)\, \bs{e}_x \, \nabla \cdot\bs{\tau}.
	\end{align}
The first integral, which reads more explicitly
	\begin{align}
		I_p &\equiv \frac{1}{L_x L_y} \int\limits_{-L_x/2}^{L_x/2} \!\!\!\!\!dx\, \int\limits_{-L_y/2}^{L_y/2} \!\!\!\!\!dy\,\, \text{sign}(x)\, \bs{e}_x \, (- \nabla p(x,y)), 
	\end{align}
can be evaluated explicitly, if we assume the local hydrostatic pressure \mbox{$p(\bs{r})$} to be only $x$-dependent, that is \mbox{$p(\bs{r}) = p(x)$}.
This is reasonable for a modulated liquid phase where $\rho(\bs{r}) = \rho(x)$, and it is consistent with our effective density profile ansatz [see Eq.~\eqref{Eq_effective_density_ansatz}].
We then obtain $\tilde{I}_p \equiv I_p L_x$ as
	\begin{align}
	 \tilde{I}_p &= (-1) \left[p\left(-\frac{L_x}{2}\right) + p\left(\frac{L_x}{2}\right)  - 2 p\left(0\right)  \right]
	 \label{Eq_Integrated_pressure_gradient}
	\end{align}
(where we have used that $\bs{e}_x \nabla p(x)= \partial_x p(x)$, and the $y$-integration cancels with $L_y$).
The pressure values entering Eq.~\eqref{Eq_Integrated_pressure_gradient} can be directly calculated by inserting the approximate density profile [see Eq.~\eqref{Eq_effective_density_ansatz}] into Eq.~\eqref{Eq_local_hydrostatic_pressure_mean_field_approximation}.
For later usage, we note that the resulting local pressure is no longer a piecewise constant function, contrary to our ansatz for the density profile.
We also note that taking the limit $L_x \searrow L_c$ from above (as before) yields
	\begin{align}
		p\left(\pm \frac{L_x}{2}\right)  = 0
	\end{align}
(with $x = \pm L_x/2$ being outside the rectangular box).
With this, we obtain
	\begin{align}
		 \tilde{I}_p &= 2 p(0) \label{Eq_Integrated_pressure_gradient_simplified}\\ \notag
				&= 2 \beta^{-1} \makeEffectiveAverageDensitySymbol  \\\notag
				&\quad\quad+ \makeEffectiveAverageDensitySymbol^2   \sum_{j = -N_r}^{N_r} \int \limits_{j L_s - L_c/2}^{j L_s + L_c/2} \!\!\!\!\!\!\!\!\!\!dx'\,\,\,\, \int \limits_{-\infty}^{\infty} \!\!\!dy'\, V(-x', y')	
	\end{align}

\begin{figure*}
    \centering	
	\includegraphics[width=0.95\textwidth]{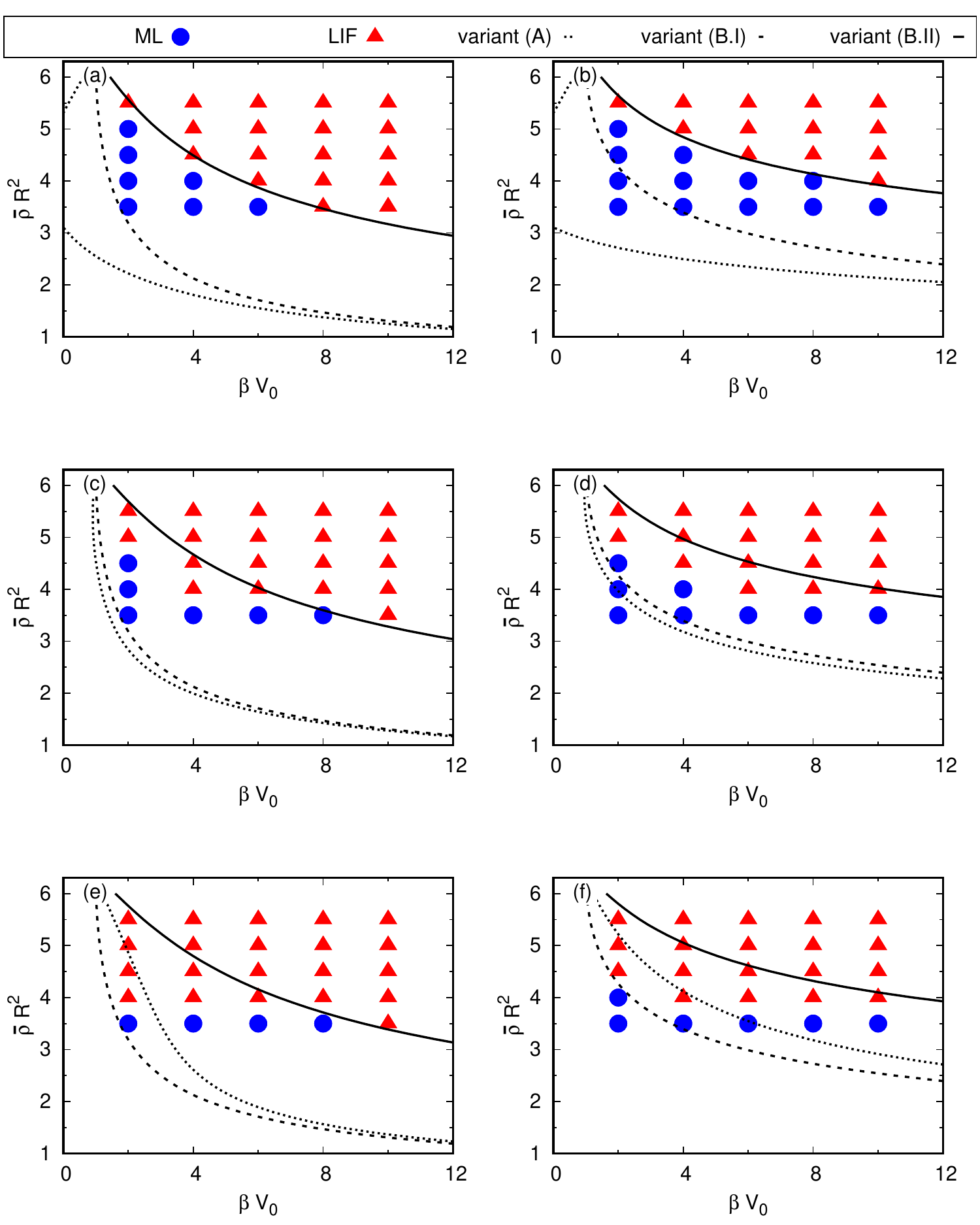} 
    \caption{Theoretical predictions from variants (A), (B.I), and (B.II) for the onset of LIF as compared to DFT calculations for the cosine substrate (a,c,e) and the Gaussian substrate (b,d,f), respectively.
    The substrate periodicity is $L_s/R = 1.8$ (top), $L_s/R = 1.6$ (center), and $L_s/R = 1.2$ (bottom).}
    \label{Fig_parameterScan_and_different_variants_of_prediction}
\end{figure*}

For $I_{\bs{\tau}}$, we proceed by inserting the effective density ansatz [Eq.~\eqref{Eq_effective_density_ansatz}] into the expression for $\divergence \bs{\tau}$ [see Eq.~\eqref{Eq_divergence_of_deviatoric_stress_tensor_mean_field_approximation}].
Multiplying by $\text{sign}(x) \bs{e}_x$ and averaging yields (in the limit $L_x  \searrow L_c$)	
	\begin{align}
		&\tilde{I}_{\bs{\tau}} = I_{\bs{\tau}} L_x = -\frac{(\bar{\rho}_{\text{eff}})^2}{2} \sum_{j = -N_r}^{N_r} \notag\\
	&\hspace{-0.75cm}\left[
	\,\,\, \int \limits_{j L_s -L_c/2}^{j L_s +L_c/2}\!\! \!\! \!\!\!\!\!dx'\, \int \limits_{-\infty}^{\infty}\!\! \!dy'\, \left(V\left(-\frac{L_c}{2} - x', y'\right) + V\left(\frac{L_c}{2} - x', y'\right) \right)  \right. \notag\\
	+	& \left. \int \limits_{-L_c/2}^{L_c/2} \!\! \!\! \!\!dx\, 
	\int \limits_{j L_s - L_c/2}^{j L_s + L_c/2}\!\! \!\! \!\!dx'\, 
	\int \limits_{-\infty}^{\infty}\!\! \!\!dy'\, 
	\text{sign}(x) \partial_x V(x-x', y') \right]. 
	\label{Eq_integrated_deviatoric_stress_tensor_MF}
	\end{align}
Combining the expressions \eqref{Eq_Integrated_pressure_gradient_simplified}, \eqref{Eq_integrated_deviatoric_stress_tensor_MF} and \eqref{Eq_averaged_external_stress_within_effective_density_ansatz}, and identifying $\makeEffectiveAverageDensitySymbol$ with  the bulk stability threshold (see Sec.~\ref{SEC:Prediction_variant_A}), we obtain again a prescription for the onset of LIF.
We call this variant (B.I).
Corresponding results are shown in Fig.~\ref{Fig_parameterScan_and_different_variants_of_prediction}.
Comparing the prediction of variant (B.I) to the DFT data, we see that it performs somewhat better than variant~(A).
However, it is still far from satisfactory.

Finally, we introduce a further variant which we refer to as (B.II). 
This is a modification of variant (B.I), where the pressure profile $p(\bs{r})$ entering Eq.~\eqref{Eq_Integrated_pressure_gradient} is not calculated through Eq.~\eqref{Eq_local_hydrostatic_pressure_mean_field_approximation}; rather we make an \textit{ansatz} for the pressure profile.
Specifically, we set
\begin{alignat}{2}
p(x)  	&=p(\bar{\rho}_{\text{eff}} )\, \text{rect}\left( \frac{x}{L_c} \right),
\label{Eq_effective_pressure_profile_ansatz}
\end{alignat}
where $p(\bar{\rho}_{\text{eff}} )$ is the pressure of a \textit{bulk} system of constant density $\makeEffectiveAverageDensitySymbol$.
This ansatz is motivated by two arguments.
(i)~The first one is consistency:
It seems reasonable that a piece-wise constant density profile [see Eq.~\eqref{Eq_effective_density_ansatz}] is accompanied by a piece-wise constant local hydrostatic pressure.
(ii)~The local hydrostatic pressure in a region of constant density should correspond to the bulk pressure at this density. 
With these assumptions Eq.~\eqref{Eq_Integrated_pressure_gradient} yields 
	\begin{align}
		 \tilde{I}_p &= 2 p(\makeEffectiveAverageDensitySymbol) \label{Eq_Integrated_pressure_gradient_with_effective_pressure_profile_ansatz} \\ \notag
		 &= 2 \beta^{-1} \makeEffectiveAverageDensitySymbol \\ \notag
		 &\quad\quad + \makeEffectiveAverageDensitySymbol^2 \int\limits_{-\infty}^{\infty} dx' \int\limits_{-\infty}^{\infty}dy' V(x',y').
	\end{align}
For $I_{\tau}$, we proceed as in variant (B.I).
Comparing the prediction for the onset of LIF from variant (B.II) to the DFT data, we find that there is now very good agreement in the case \mbox{$L_s/R=1.8$}.
Further, for \mbox{$L_s/R=1.6$}, the prediction is considerably improved.
We also performed calculations for substrate periodicities $L_s / R$ = 1.4, 1.2, and 1.0.
The strongest deviations between DFT data and the prediction occurs at $L_s / R = 1.2$ (as discussed in Sec.~\ref{SUBSEC:Phase_diagrams}).
We attribute this to the fact that $L_s / R = 1.2$ is the optimal periodicity for the GEM-4 potential (see Sec.~\ref{SUBSEC:Phase_diagrams}).
Therefore, and since our prediction is based on a sufficient criterion (see discussion at the end of Sec.~\ref{SEC:Prediction_variant_A}), the onset of freezing is underestimated at $L_s / R = 1.2$.

\subsection{Discussion of variant (B.II) from a physical perspective}

Interestingly, variant (B.II) not only yields the best prediction of the LIF transition; it also allows for an interpretation in terms of an effective-fluid picture.
To see this, we explicitly consider Eq.~\eqref{Eq_averaged_stress} in variant~(B.II) by combining Eqs.~\eqref{Eq_Integrated_pressure_gradient_with_effective_pressure_profile_ansatz}, \eqref{Eq_integrated_deviatoric_stress_tensor_MF}, and \eqref{Eq_averaged_external_stress_within_effective_density_ansatz}.
This yields
	\begin{align}
		\label{Eq_averaged_stress_within_effective_density_ansatz_variant3}
	2 \, p(\bar{\rho}_{\text{eff}})    + \tilde{I}_{\bs{\tau}}( \bar{\rho}_{\text{eff}}, L_c  )
	&= 2\, \bar{\rho}_{\text{eff}}  V_{\text{ext}}\left(\frac{L_c}{2} \right).
	\end{align}
Equation ~\eqref{Eq_averaged_stress_within_effective_density_ansatz_variant3} may be viewed as a pressure balance in an effective \textit{bulk} fluid of density $\bar{\rho}_{\text{eff}}$.
The inhomogeneity caused by the external potential $V_{\text{ext}}$ is reflected only indirectly by the appearance of the contribution from the deviatoric stress $\tilde{I}_{\bs{\tau}}$ (which would be zero in a true bulk fluid).
Further, it is instructive to rewrite Eq.~\eqref{Eq_averaged_stress_within_effective_density_ansatz_variant3} into
\begin{align}
	\label{Eq_EOS_form_averaged_stress_within_effective_density_ansatz}
 \frac{\beta \,  p(\bar{\rho}_{\text{eff}})  }{\bar{\rho}_{\text{eff}}} + \frac{ \beta \, \tilde{I}_{\bs{\tau}}( \bar{\rho}_{\text{eff}}, L_c  )}{2 \bar{\rho}_{\text{eff}}} = \beta V_{\text{ext}}\left(\frac{L_c}{2} \right),
\end{align}
where the first term, $Z = \beta p(\bar{\rho}_{\text{eff}}) / \bar{\rho}_{\text{eff}}$, may be considered as the compressibility factor of the effective bulk fluid.

Equation~\eqref{Eq_EOS_form_averaged_stress_within_effective_density_ansatz} is indeed a central result of our work.
To acknowledge this, we remark that it can be derived without explicitly assuming any particular form for the particle interactions and the correlation functions.
These are encapsulated within the compressibility factor $Z$ and the correction term due to inhomogeneity $\tilde{I}_{\bs{\tau}}$.
Due to its rather general structure, it is possible to use Eq.~\eqref{Eq_EOS_form_averaged_stress_within_effective_density_ansatz} also for other systems beyond the ultra-soft fluid considered here.
This will be demonstrated in a forthcoming work\cite{Kraft_futurePaper}.

\section{Conclusion\label{SEC:Conclusion}}

In this study we used classical density functional theory to study the freezing behaviour of a 2D system of ultra-soft particles in the presence of two variants of 1D periodic potentials. Our results from a free minimization of the (mean-field) density functional clearly show that an ensemble of soft particles interacting via a GEM-4 potential displays the phenomenon of LIF, although the repulsive interaction is bound and therefore differs significantly from the previously investigated cases. By studying different variants of the external potential, we have, moreover, found evidence that the mechanism driving the transition is
the increase of the density inside the inner ("confining") zone of the potential wells, $\bar{\rho}_{\text{eff}}$. Indeed, quite independent of the details of the potential, we find LIF to occur when $\bar{\rho}_{\text{eff}}$ reaches a certain threshold value.
This finding provides a somehow alternative view to the conventional LIF scenario, where the key parameter is the amplitude of the potential, $V_0$. 
In our picture, $V_0$ (or the width of the wells for the Gaussian substrate) is rather considered as an external parameter whose variation may drive an increase of $\bar{\rho}_{\text{eff}}$, thereby triggering LIF.

Using this picture, we have developed a new strategy to predict theoretically the onset of LIF. Our approach foots on an integrated version of the (exact) pressure balance equation,
which we evaluate in the modulated liquid phase. To this end, we use a rectangular parametrization of the density profile (involving $\bar{\rho}_{\text{eff}}$ and the confining length, $L_c$). Approximating the threshold density by the {\em bulk} stability threshold allows us to obtain the {\em external} parameters, such as $V_0$, related to the LIF transition.
Exploiting this strategy for the soft system at hand, which is well described by a mean-field approximation, we find surprisingly good agreement with the data from free minimization. We note that the accuracy
depends on the route of evaluation of the terms in the resulting pressure-balance equation. This strongly reminds of the phenomenon of thermodynamic inconsistency well known in liquid state theory\cite{HansenMcDonald}.

Clearly, the present approach markedly differs from the more established DFT approaches to freezing such as RY theory\cite{RamakrishnanYussouff1979, Ramakrishnan1982}. An advantage here is that we do not need the full direct correlation function of the reference state, from which the freezing occurs, as an input. The present approach rather involves the bulk pressure (as function of density), as well as the stability threshold of the bulk system. 
In this sense its structure is somewhat simpler. 

Still, it seems fair to discuss some open questions. One of these questions concerns the simplified (rectangular) {\em ansatz} for the density profile in the modulated liquid phase.
In the actual physical system with finite potential barriers, the individual particles are clearly {\em not} completely confined within a region $L_c$ around the minimum. Moreover, the density inside the wells is certainly not constant in $x$-direction. One should note, however, that our approach involves an {\em integrated} exact equation, such that density inhomogeneities do not contribute explicitly.
Rather, one may consider the rectangular profile as a convenient starting point which is consistent with the usage of the mean value theorem of integration. 

A further important question is to which extent the present approach could be transferred to other model systems.
To this end we recall that the final form of the integrated pressure balance equation
has a quite general,  "bulk-like", structure with intuitive physical interpretation: Apart from the parameters of the external potential, the equation involves the compressibility factor (that is, the isotropic pressure) of a bulk fluid of density $\bar{\rho}_{\text{eff}}$ and a deviatoric contribution stemming from the inhomogeneities. 
Provided that there is a theoretical prediction or numerical data for these quantities, one could use the equation for other model fluids as well. 
The application to the hard disk system will be discussed in a forthcoming work.


\appendix

\section{Technical details of the DFT calculations\label{SEC:Technical details of the DFT calculations}}

We use the standard iteration method for the free minimization of the grand potential functional (as described, e.g. in Ref.~\onlinecite{Hughes2014}).
Since LIF corresponds to an induced liquid-solid phase transition of an otherwise stable liquid phase (in the absence of an external potential), we use a homogeneous density profile (plus noise) as initial condition.
We studied different system sizes with quadratic aspect ratio $L_x = L_y \geq 20 R$ and with rectangular aspect ratio $L_x/L_y = \sqrt{3}/{2}$ and found no differences in the arising phases in the phase diagrams.
For the scan of the phase diagrams, the discretization $dx = dy = 0.05 R$ was used. 
This relatively large value was chosen due to the large number of calculations and constraints in computational time. 
The more detailed calculations were performed with higher numerical resolution $dx = dy = 0.005 R$.
We note that points close to the phase boundaries (see, e.g. Fig.~\ref{Fig_CompareParameterScanWithPrediction_L_s=1.6}) can be subject to convergence problems in the sense that it becomes difficult to identify the locked floating solid phase against the modulated liquid. 
These problems are absent deep in the frozen phase.

\section{Details of the effective density profile parametrization \label{SEC:Details_of_the_effective_density_profile_parametrization}}

In this Appendix, we discuss the details of the effective density ansatz [see Eq.~\eqref{Eq_effective_density_ansatz}] for the density profile~$\rho(x)$.
The periodic repetition (with the substrate periodicity $L_s$) of the ansatz yields
\begin{align}
	\rho_p(x) 	= \sum_{j= -N_r}^{N_r} \rho(x + j L_s),
\end{align}
where $N_r$ is the number of repetitions (and thus the number of neighbouring minima which are taken into account for particle interactions across adjacent minima).
We also state their gradients, which read
\begin{align}
\nabla \rho(x) 	= \bs{e}_x \bar{\rho}_{\text{eff}} \left[ \delta\left(x + \frac{L_c}{2} \right)  -  \delta\left( x - \frac{L_c}{2} \right)  \right],
\label{Eq_effective_density_ansatz_derivative}
\end{align}
and with periodic repetition 
\begin{align}
 \nabla \rho_p(x) 	&= \sum_{j= -N_r}^{N_r} \nabla \rho(x + j L_s) \\
				&= \bs{e}_x \bar{\rho}_{\text{eff}} \sum_{j= -N_r}^{N_r}  \left[ \delta\left(x + j L_s + \frac{L_c}{2} \right)  \right.  \notag\\
				&\left.\hspace{2.2cm}  -  \delta\left( x  + j L_s - \frac{L_c}{2} \right)  \right].
\end{align}
In our actual calculations, we set the number of repetitions (i.e. the number of neighbouring minima that are taken into account) to $N_r=1$.
We did not observe a (numerically significant) quantitative difference for larger $N_r$.

\section{Onset of LIF for varying $V_0$ \label{Onset_of_LIF_for_varying_V_0}}

In this Appendix, we explicitly show that when the potential amplitude $V_0$ is varied to induce the LIF transition, we can explicitly solve for the value of $V_0$ for the onset of LIF for all variants [see variants (A), (B.I), and (B.II) in Sec.~\ref{SEC:Explicit_calculations_for_our_model_system}].
Using Eq.~\eqref{Eq_redistribution_of_particles}, the required confining length $L_c$ to enforce the increase of the density from the average system density $\bar{\rho}$ to a given value of the effective average density~$\bar{\rho}_{\text{eff}}$ is given by
\begin{align}
 L_c &=  \frac{\bar{\rho}}{\bar{\rho}_{\text{eff}}}   L_s .
\end{align}
For the external potential in a typical LIF transition, we can explicitly factor out the potential amplitude $V_0$ such that \mbox{$V_{\text{ext}} (x)= V_0 \cdot \tilde{V}_{\text{ext}} (x)$}.
Using Eqs.~\eqref{Eq_averaged_stress}, \eqref{Eq_averaged_external_stress_within_effective_density_ansatz}, and \eqref{Eq_Integration_of_div_stress_tensor}, the required potential amplitude~$V_0$ to enforce the relocation of particles from $L_s$ to $L_c$ (thus causing an increase from $\bar{\rho}$ to $\bar{\rho}_{\text{eff}}$) within our prediction occurs at 
\begin{align}
	V_0 &= \frac{ \tilde{I}_{\bs{\sigma}}}{2 \makeEffectiveAverageDensitySymbol \tilde{V}_{\text{ext}}\left(\frac{L_c}{2}\right)}.
\end{align}
The theoretical prediction consists in prescribing a threshold value $\bar{\rho}_{\text{eff}} = \bar{\rho}_{\text{eff,c}}$ which the effective average density has to exceed at the LIF phase transition.
The above $V_0$ then yields the required potential amplitude for the onset of LIF.

\bibliography{myReferences}{}

\end{document}